\documentclass[english]{emulateapj}
\usepackage[T1]{fontenc}
\usepackage{epstopdf}
\usepackage[latin9]{inputenc}
\usepackage{verbatim}
\usepackage{rotfloat}
\usepackage{textcomp}
\usepackage{graphicx}
\usepackage{amssymb}
\usepackage{esint}

\DeclareRobustCommand{\greektext}{%
  \fontencoding{LGR}\selectfont\def\encodingdefault{LGR}}
\DeclareRobustCommand{\textgreek}[1]{\leavevmode{\greektext #1}}
\DeclareFontEncoding{LGR}{}{}

\providecommand{\tabularnewline}{\\}

\usepackage{babel}

\begin{document}

\title{The Curious Case of Glass I: High Ionization and Variability of Different Types}

\author{Andrew J. Kruger}
\affil{Department of Physical Science, Wilbur Wright College, 4300 N. Narragansett Ave, Chicago, IL 60634}

\author{Matthew J. Richter}
\affil{Department of Physics, University of California at Davis, One Shields Ave, Davis, CA 95616, USA}

\author{John S. Carr}
\affil{Remote Sensing Division, Naval Research Laboratory, Code 7210, Washington, DC 20375, USA}

\author{Joan R. Najita}
\affil{National Optical Astronomy Observatory, Tucson, AZ 85719, USA }

\author{Margaret M. Moerchen}
\affil{European Southern Observatory, Alonso de C\'{o}rdova 3107, Santiago, Chile}
\affil{Leiden Observatory, PO Box 9513, 2300 RA Leiden, The Netherlands}

\author{Greg W. Doppmann}
\affil{W.M. Keck Observatory, 65-1120 Mamalahoa Hwy, Kamuela, HI 96743}

\and
\author{Andreas Seifahrt}
\affil{Department of Astronomy and Astrophysics, University of Chicago, 5640 S. Ellis Ave, Chicago, IL 60637, USA}

\begin{abstract}

Our {\it Spitzer} IRS observation of the infrared companion Glass Ib revealed fine structure emission with high ionization ([NeIII]/[NeII]=2.1 and [SIV]/[SIII]=0.6) that indicates the gas is likely illuminated by hard radiation.  While models suggest extreme ultraviolet radiation could be present in T Tauri stars (Hollenbach \& Gorti 2009 and references therein), this is the first detection of [SIV] and such a high [NeIII]/[NeII] ratio in a young star.  We also find that Glass Ib displays the molecules HCN, CO$_2$, and H$_2$O in emission.  Here we investigate the Glass I binary system and consider possible mechanisms that may have caused the high ionization, whether from an outflow or disk irradiation.  We also model the spectral energy distributions of Glass Ia and Ib to test if the system is a young member of the Chameleon I star-forming region, and consider other possible classifications for the system.  We find Glass Ib is highly variable, showing changes in continuum strength and emission features at optical, near-infrared, and mid-infrared wavelengths.  The optical light curve indicates that a central stellar component in Glass Ib became entirely visible for 2.5 years beginning in mid-2002, and that possibly displayed periodic variability with repeated, short-period dimming during that time.  As the fine structure emission was not detected in observations before or after our {\it Spitzer} IRS observation, we explore whether the variable nature of Glass Ib is related to the gas being highly ionized, possibly due to variable accretion or an X-ray flare.

\end{abstract}

\keywords{stars: pre-main-sequence -- infrared: stars -- planetary systems: protoplanetary disks -- accretion, accretion disks -- stars: individual (Glass I)}

\section{Introduction }

One of the defining characteristics of young solar mass stars, T Tauri stars, is they show variability (Joy 1945).  These objects are known to be variable on short time-scales at optical, near-infrared and mid-infrared wavelengths.  These variations may be due to changes in accretion, the central star (e.g. photospheric spots), or even in the disk structure.  For a complete overview of T Tauri Stars, refer to reviews by M\'enard \& Bertout (1999) and Petrov (2003).

Although originally classified as a field star, Chelli et al.~(1988) found Glass I to be a binary, consisting of a weak-lined T Tauri K-star and a nearby companion showing strong infrared excess.  Koresko et al.~(1997) later included Glass Ib in a list of classical infrared companions (IRC).   IRCs are close companions to young stellar objects (YSO) that dominate at infrared wavelengths but have such high optical extinction they are often invisible in visible wavelengths.  We were granted time with {\it Spitzer} IRS to observe edge-on circumstellar disks and classical infrared companions; the Glass I binary system was among the targets in the study.   We were surprised to find the Glass I binary showed fine structure emission with higher ionization ratios than previously seen toward YSOs.

\begin{figure*}
\begin{center}
\includegraphics[trim=82 80 95 95,clip,angle=90,scale=0.65]{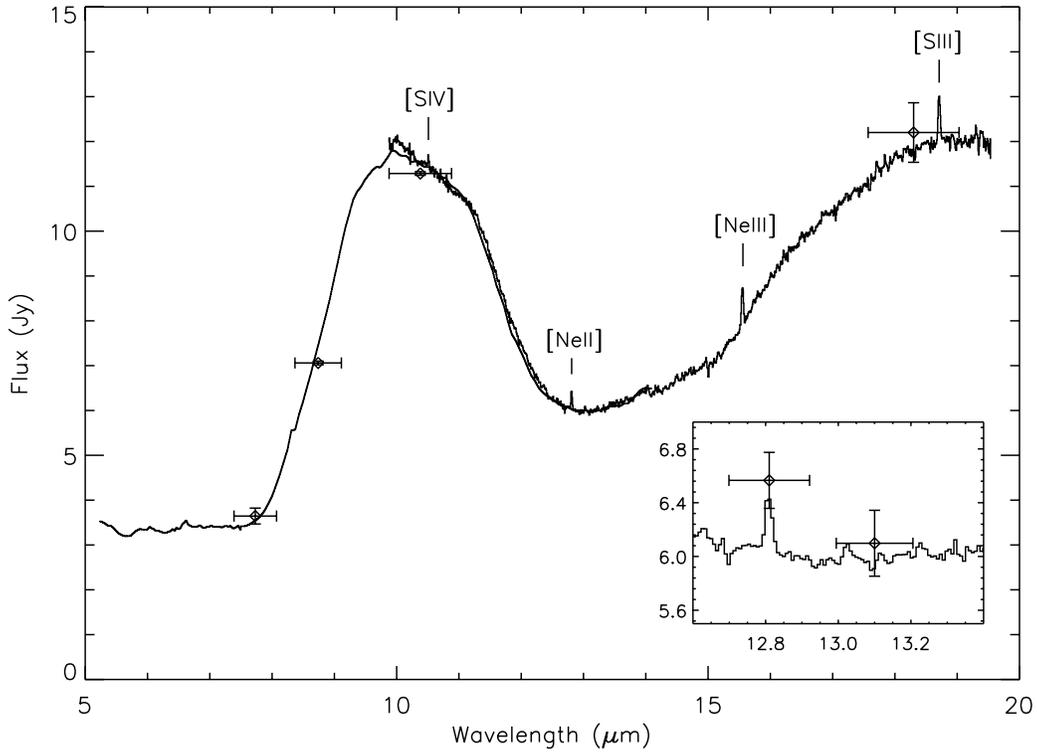}

\caption{The {\it Spitzer} IRS short-low and short-high spectra of the Glass I binary system. Overplotted
are the photometric fluxes for Glass Ib found using T-ReCS, with the
error widths showing the FWHM of the filters. The inset shows the {[}NeII{]} and {[}NeII{]}-continuum
filters.}
\end{center}
\end{figure*}

Extreme UV and X-ray irradiation may play an important role in planet formation as they may be responsible for mechanisms such as chemical processing and disk dissipation.  The study of such irradiation in disks is thus important for models of the synthesis of organic molecules and ultimately planets themselves.  Disk models have included photoevaporation by EUV radiation to explain gas-poor giant planets (i.e. Uranus, Neptune), T Tauri disk dispersal rates, the presence of inner disk holes, and planetary migration (Hollenbach \& Gorti 2009 and references therein).  Extreme UV and X-ray irradiation also heat and ionize the disk surface.  This upper layer may be accreted onto the central star as the result of magnetorotational instability (Balbus \& Hawley 1991) while planets form in the disk midplane, which is shielded from the stellar radiation by the surface layers.  Models also show the gradients in temperature and ionization can drive molecular evolution in disks (Aikawa \& Herbst 1999).

The origin of the EUV and X-ray radiation is not fully understood.  The radiation may come from the accretion shocks of infalling matter or from strong magnetic activity in the central stellar object.  However, the EUV photons have a short penetration depth, so if this radiation is being produced by accreting material, it can be absorbed by the accreting material before it reaches the disk.  The short penetration depth of EUV photons also makes it very difficult to directly detect this radiation because it is also greatly absorbed by interstellar hydrogen (Drake 1999).  In order to investigate the radiation environment in disks, fine structure emission lines originating from the ionized disk surface, such as [NeII] and [NeIII], have been proposed as diagnostics (Glassgold et al.~2007; Gorti \& Hollenbach 2008).  The detection of [NeII] emission in YSOs has become common with {\it Spitzer} IRS (e.g. Lahuis et al.~2010).  This line may be due to just X-ray ionization of the gas (Glassgold et al.~2007), supported by Pascucci et al.~(2007) finding a correlation between X-ray luminosity and [NeII] line strength.  However, EUV photons may play a role in creating the [NeII] emission if the accretion rate is $<10^{-8}$ M$_{\odot}$ yr$^{-1}$, which would allow the EUV photons to reach the disk (Hartmann et al.~1998).  This was supported by Espaillat et al.~(2007) who did not find a correlation between [NeII] line strength and X-ray luminosity, but rather with the mass accretion rate of the disk.  G\"udel et al.~(2010) found with a large sample of YSOs showing [NeII] emission that it is likely the case that both disk accretion and X-ray luminosity play roles in the strength of this line.  G\"udel et al.~also found that systems with jets show the strongest [NeII] emission, showing this to be a separate mechanism that can create this line.  The velocity profiles of [NeII] emission from some YSOs show high velocity shifts, showing they likely originating from the outflow shocks (Shang et al.~2010 and references therein), so it is important to distinguish between emission that is due to an outflow or an irradiated disk.

While the detection of [NeII] emission is common, the detection of highly ionized atomic species such as [NeIII] and [SIII] are rare.  The highest [NeIII]/[NeII] ratio that has been found to date is $\sim$1 toward SZ Cha, while other ratios are typically found to be <0.5 in systems showing [NeIII] emission (Espaillat et al.~2012).  This may be because the recombination rate for Ne$^{2+}$ is much higher than Ne$^{+}$.  If the gas is being ionized by X-ray radiation, the photons will penetrate deeper into the disk where there are more free electrons to recombine with the Ne$^{2+}$ atoms, resulting in a low [NeIII]/[NeII] ratio.  Extreme UV photons would not penetrate as deep into the disk, resulting in a heated disk surface that can create a higher [NeIII]/[NeII] ratio (Hollenbach \& Gorti 2009).  In the case of the emission lines originating from an outflow, Shang et al.~(2010) predict with models that both [NeII] and [NeIII] emission can arise from a jet, and high ionization states such as [NeIII] have been detected in the Herbig-Haro outflow seen toward GGD 37 (Green et al.~2011).

Here we investigate the Glass I system in an effort to understand the source of the photoionization.  We explore whether the highly variable nature of this object can provide clues to the ionizing source, and even consider if Glass Ib is truly a young star.  We describe the observations and their results in section 2, discuss the results in section 3, and finish with a summary of our findings in section 4.

\begin{table*}
\begin{center}

\caption{Detailed List of Observations Taken and Archival Data Used in this Work.}

\begin{tabular}{cccccl}
\hline 
Instrument & Date & Mode & Setting & Total Int. Time (s) & Notes \tabularnewline
\hline
\hline 
WFI & 2003 December 23 & Image & 676 nm & 900 & Glass Ib saturated image\tabularnewline 
WFI & 2004 April 13-24 & Image & 830 nm & 5400 & Glass Ib saturated image\tabularnewline
FORS & 2005 April 1 & Image & U band & 3330 & $\Delta$U$^{a}$=0.31$\pm$0.05 mag\tabularnewline 
& & & B band & 960 & $\Delta$B$^{a}$=0.56$\pm$0.05 mag \tabularnewline 
& & & V band & 540 & $\Delta$V$^{a}$=0.58$\pm$0.05 mag\tabularnewline 
& & & R band & 120 &  $\Delta$R$^{a}$=0.55$\pm$0.05 mag \tabularnewline
Spitzer IRS & 2006 March 7 & Spectrum & SH$^{b}$ & 6 & No fine structure emission\tabularnewline
Spitzer IRS & 2008 August 16 & Spectrum & SL1, SL2 & 151$^{c}$ & \tabularnewline
& & & SH & 24 & Fine structure emission \tabularnewline
T-ReCS & 2009 June 8 & Image & Si-1, 7.73 \textgreek{m}m & 697 & F$^{d}$=3.65$\pm$0.2 Jy\tabularnewline
& & & Si-2, 8.74 \textgreek{m}m & 694 & F$^{d}$=7.06$\pm$0.04 Jy\tabularnewline
& & & Si-4, 10.38 \textgreek{m}m & 696 & F$^{d}$=11.3$\pm$0.04 Jy\tabularnewline
& & & [NeII], 12.81 \textgreek{m}m & 807 & F$^{d}$=6.57$\pm$0.2 Jy\tabularnewline
& & & [NeII] cont, 13.10 \textgreek{m}m & 805 & F$^{d}$=6.10$\pm$0.2 Jy\tabularnewline
& & & Qa, 18.3 \textgreek{m}m & 892 & F$^{d}$=12.2$\pm$0.7 Jy\tabularnewline
X-Shooter & 2010 January 20 & Spectrum & UVB & 180 \tabularnewline
& & & VIS & 160 \tabularnewline
& & & nIR & 180 \tabularnewline
VISIR & 2010 January 26, 27 & Spectrum & 12.81 \textgreek{m}m & 1210 & No [NeII] emission\tabularnewline

\hline
\\
\multicolumn{6}{l}{$^{a}$M(Glass Ib) - M(Glass Ia).  $^{b}$SH data obtained from {\it Spitzer} Heritage Archive, PI: Houck, AOR: 12697088. }\tabularnewline

\multicolumn{6}{l}{$^{c}$Integration time for each setting. $^{d}$Measured flux is for Glass Ib.  }\tabularnewline

\end{tabular}

\end{center}



\end{table*}

\section{Observations and Results}\label{obs_res}

\subsection{{\it Spitzer} IRS Spectrum}

The Glass I binary system was part of our {\it Spitzer} Cycle 5 General Observers campaign,
Program ID 50152, to use the Infrared Spectrograph (IRS; Houck et
al. 2004) in short-low and short-high (SL and SH) stare mode. 
The SL module has a slitwidth of 3.6$^{\prime\prime}$, plate scale of 1.8$^{\prime\prime}$ pixel$^{-1}$, and covers the spectral range 5.2-14.5 \textgreek{m}m with resolution R=60-127, while the SH module has a slitwidth of 4.7$^{\prime\prime}$, plate scale of 2.3$^{\prime\prime}$ pixel$^{-1}$, and covers 9.9-19.6 \textgreek{m}m with R=600.  The observations were designed to search for weak absorption features and 
so followed the observing technique described in Carr \& Najita
(2008). We shortened the exposure time of each data frame to 6 seconds
and increased the number of integrations to verify repeatability of
the observations. Each setting (SL1, SL2, and SH) had twelve on-source
data frames at each nod position, accompanied with half as many off-source
integrations of the blank sky for hot pixel subtraction and to remove background emission.
The observation (AOR: 25680640) was taken on 2008 August 16 (see Table 1) at 
$11^{h}08^{m}15.8^{s}\mbox{ }-77^{\circ}33^{\prime}53^{\prime\prime}$.
The coordinates were chosen to center the slit on Glass Ib, calculated
by using the position of Glass Ia from SIMBAD and the offsets found
by Chelli et al.~(1988). The observation used the nominal pointing
accuracy of $0.1^{\prime\prime}$.

The low resolution SL spectra were reduced using optimal point source
extraction in the {\it Spitzer} IRS Custom Extractor (SPICE; version 2.2).
The high resolution SH spectra were reduced following the reduction
technique described in Carr \& Najita (2008) using custom IDL routines
along with standard spectral reduction routines in IRAF. 
After identifying and removing bad pixels found using the images of the nearby blank sky,
we used the \emph{apall} routine in IRAF to extract the raw spectra.
We also extracted the spectra for observations of HR 6688, spectral type K2 III,which we
used  for standard star division to remove fringing and for flux calibration.
Because fringing is dependent on the target position in the slit, we used
the standard star observation that resulted in the least amount of
fringing at shorter wavelengths. We then used IRSFRINGE in IDL to remove the residual fringing.

\begin{figure*}
\begin{center}
\includegraphics[trim=305 80 90 90,clip,angle=90,scale=0.6]{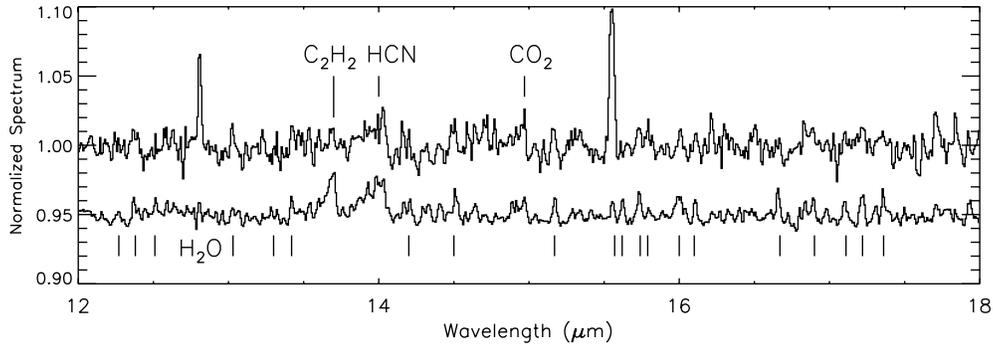}

\caption{The normalized {\it Spitzer} IRS spectrum of the Glass I system (top spectrum) compared to the spectrum for DoAr 24E (bottom spectrum), an object from our {\it Spitzer} IRS campaign that was also observed by Pontoppidan et al.~(2010). We indicate the location of the HCN, CO$_2$, and C$_2$H$_2$  Q-branches; while HCN and CO$_2$ are detected in Glass I, C$_2$H$_2$  emission is not clearly present.  The marked H$_2$O emission lines are from Pontoppidan et al.~(2010). 
}
\end{center}
\end{figure*}

\begin{table*}
\begin{center}
\caption{Fine structure line intensities as measured with {\it Spitzer} IRS.}

\begin{tabular}{ccccc}
\hline 
 & & \textgreek{l} & Line Intensity$^{a}$ (2008) & 5$\sigma$ Upper Limits (2006)\tabularnewline
Ion & Transition & (\textgreek{m}m) & ($10{}^{-14}\mbox{ erg cm}^{-2}\mbox{ s}^{-1}$) & ($10{}^{-14}\mbox{ erg cm}^{-2}\mbox{ s}^{-1}$)\tabularnewline
\hline
\hline 
{[}S IV{]} & $^{2}P_{3/2}-^{2}P_{1/2}$ & 10.51 & 4.70$\pm$1.4 & 0.95\tabularnewline
{[}Ne II{]} & $^{2}P_{1/2}-^{2}P_{3/2}$ & 12.81 & 3.59$\pm$0.7 & 0.84\tabularnewline
{[}Ne III{]} & $^{3}P_{1}-^{3}P_{2}$ & 15.56 & 7.53$\pm$0.9 & 0.79\tabularnewline
{[}S III{]} & $^{3}P_{2}-^{3}P_{1}$ & 18.71 & 7.41$\pm$1.2 & 0.56\tabularnewline
\hline
\\
\multicolumn{5}{l}{$^{a}$Line intensities are not extinction corrected.}
\end{tabular}
\end{center}
\end{table*}

The final IRS spectrum, shown in Figure 1, is almost entirely from Glass Ib. The flux contribution from Glass Ia is likely negligible in this spectrum as shown by our T-ReCS images discussed below.  The most obvious feature in the {\it Spitzer} IRS spectrum is the  silicate emission that has been seen toward Glass Ib before (Meeus et al.~2003).   The spectrum also shows clear {[}NeII{]}, {[}NeIII{]}, {[}SIII{]}, and {[}SIV{]} emission, indicated in Figure 1.  The integrated fluxes of these lines, without correction for extinction, are in Table 2.  While {[}NeII{]} is seen toward circumstellar disks (Lahuis et al.~2007), {[}NeIII{]} emission is unusual with most detections at low signal-to-noise (i.e. Sz 102, Lahuis et al.~2007; TW Hya, Najita et al.~2010).  In the case of Glass Ib, the [NeIII] flux is actually twice as strong as the [NeII].  Similarly, {[}SIII{]} emission has been detected toward disk sources (Lahuis et al.~2007), but not {[}SIV{]}.  A further search of the literature for the high ionization atomic line emission in other YSO systems revealed only that [NeV] was seen in association with the GGD 37 outflow (Green et al.~2011), but the [NeIII]/[NeII] was near $\sim0.2-0.5$ and there was no [SIII] or [SIV].

Because these forbidden lines are unusual in disk stars, we verified that they were present in each of the twelve individual IRS spectra we took of the Glass I binary system.  We find there are no hot pixels at the position of the lines, the emission is located at the Glass Ib continuum in each nod position, and the emission was not found in the background observations or in observations taken directly before or after Glass I.  We also investigated the point spread function (PSF) of the raw spectrum, and found the width and center of the beam of the fine structure emission lines are similar to the trend found at other wavelengths, indicating the emission is unresolved.

While emission from high ionization states like [NeIII] and [SIV] is unusual for T Tauri stars, Glass Ib also has molecular emission that is fairly typical for YSOs: HCN, CO$_2$, and H$_2$O.   Emission from these molecules has been seen before (Carr  \& Najita 2008; Carr \& Najita 2011; Salyk et all, 2008;  Pontoppidan et al.~2010), and originates in the disk atmosphere in the inner regions of the circumstellar disk.  To support the identification of molecular emission from Glass Ib, we compare the spectrum with another object observed in our IRS campaign, DoAr 24E, which shows these molecular species (Pontoppidan et al.~2010; Kruger et al.~2012) in Figure 2.  We also indicate C$_2$H$_2$, which is clearly in emission in the DoAr 24E spectrum and may also be weakly in emission in the Glass I spectrum

\subsection{T-ReCS Imaging}

To search for extended emission and to find flux ratios between binary
components, we used T-ReCS (Telesco et al.~1998) on Gemini South to
image the Glass I binary components in 6 filters ranging from 7.7-18.3 \textgreek{m}m (see Table
1). We took calibration
images of standard stars before and after the Glass I binary.  Observations were taken by cycling through all six filters once on target. 
We centered the targets
on the field of view for half the total integration time, and
placed the targets in a corner of the array for the other half of the time for frames that could be used for background subtraction.
These observations were taken 2009 June 8.

We used custom IDL routines to reduce the T-ReCS images. High winds
during the observation greatly reduced the image quality, so the images
were centered on the peak of the target flux before combining. We searched for extended emission by varying the apertures, but
found none.  
Deconvolution using \emph{lucy} in IRAF gave the same results.
In general, we only saw Glass Ib.  We were able to identify the primary, Glass Ia,
in the deconvolved images of the Si-2 and Si-4 filters 
as they had better signal-to-noise.
Using the primary detections in the Si-2 and Si-4 filters, and using the T-ReCS plate scale
of $0.09^{\prime\prime}$ pixel$^{-1}$, we find a binary separation of $2^{\prime\prime}.44\pm0.1$
with Glass Ib at a position angle $104.8^{\circ}\pm0.1$ relative to
Glass Ia. The photometric fluxes reported in Table 1 were found using a 1$^{\prime\prime}$
aperture radius, and are consistent with the {\it Spitzer} IRS continuum, as shown in Figure 1.

\subsection{VISIR Spectrum}

We used ESO's VLT Imager and Spectrometer for the mid-InfraRed (VISIR; Lagage et al.~2004) on 2010 January 26
and 27 to investigate the {[}NeII{]} emission found in the {\it Spitzer}
IRS spectrum. We used VISIR in high-resolution, longslit spectroscopy
mode with the [NeII] filter and a $0.4^{\prime\prime}$ slit width (56 AU assuming a distance of 140 pc to the system). The
detector has a plate scale of $0.127^{\prime\prime}\mbox{ pixel}^{-1}$ (18 AU),
and resolution R$\sim$25,000, making it suitable to get spatial and
spectral information on the emission. The observations were taken
in chop-nod mode with $8^{\prime\prime}$ chops and nods.  The observations
were taken at four position angles (75, 165, 255, and 345$^{\circ}$)
so we could do spectro-astrometric analysis (Bailey 1998) to get detailed
spatial structure of the emission.  We also observed HD 48915B to be used as a telluric divisor.  

The Glass Ib and standard star spectra were created using the VISIR pipeline.  After telluric division, the final spectrum surprisingly revealed no {[}NeII{]} emission.   The final spectrum, shown in Figure 3, has high signal-to-noise and no emission.  The apparent absorption feature at 12.812 $\mu$m is due to the remnant of a telluric CO$_2$ feature after telluric correction.  We created custom IDL routines to search for extended emission and found none.  If the [NeII] emission were uniformly distributed over the {\it Spitzer} IRS aperture, our VISIR observations still should have detected the emission along the slit.  To compare with the {\it Spitzer} IRS spectrum, we include Gaussian curves with a range of velocity widths typically found in YSOs (20-100 km s$^{-1}$; Sacco et al.~2012) and the same line intensity as measured with IRS.  If we assume an emission line with a width of 20 km s$^{-1}$, we find the 5-$\sigma$ upper limit of the line intensity to be $5.0\times10^{-15}$ ergs s$^{-1}$ cm$^{-2}$.  This was found by measuring feature strengths at different wavelengths across the spectrum, and setting the upper limit as the intensity that would provide a 5-$\sigma$ integrated intensity.  One possibility for why the [NeII] line was not detected is that the emission extended, possibly produced in a jet outside the field of view of the slit (see discussion in section 3.3.1).

\begin{figure}
\begin{center}
\includegraphics[trim=84 80 98 105,clip,angle=90,scale=0.41]{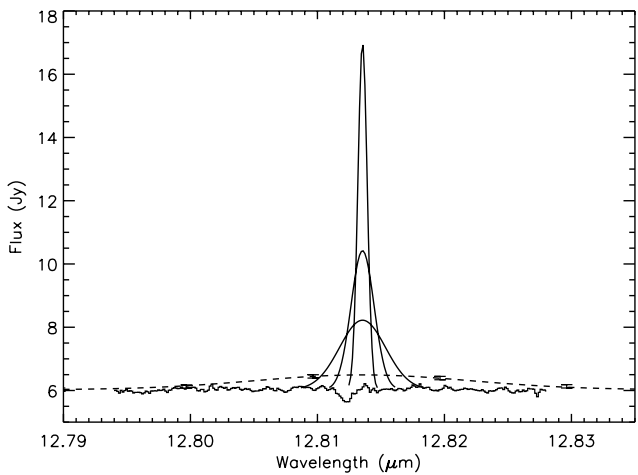}

\caption{The mid-infrared continuum as measured with VISIR on the
VLT, adjusted to be at the flux of the Spitzer spectrum, showing no {[}NeII{]} emission. Overplotted are the expected line fluxes (solid lines), assuming a Gaussian shape, if the emission was spatially unresolved and had line widths FWHM=20, 50, and 100 km s$^{-1}$ as typically found in YSOs (Sacco et al.~2012).
The {\it Spitzer}
spectrum is overplotted with spectral points indicated with error
bars, along a Gaussian fit to this feature (dashed lines) implying
a maximum FWHM of $470\mbox{ km s}^{-1}$.}
\end{center}
\end{figure}

\subsection{Optical Spectrum}

We obtained archival data of the Glass I binary observed with X-Shooter (D'Odorico et al.~2006) on the VLT on 2010 January 20
(PI: Herzceg, Prog.ID: 084.C-1095(A)). The binary was
positioned along the slit axis and a single AB nod pair
was taken with a total observing time of 180 s in UVB (340-600 nm),
160 s in VIS (560-1020 nm), and 180 s for the nIR arm (1000-2400 nm).
We used ESO's data reduction pipeline (version 1.3.0) to perform the
standard steps of data reduction (i.e. dark and flatfield correction)
and to extract rectified and wavelength calibrated 2D spectra.
We then extracted 1D spectra of the well separated components
of the Glass I binary with a custom IDL script.

No telluric standard star was observed for the Glass I system. We used the spectrum
of a flux standard (GD 71) in the same night to correct for the
instrumental throughput in UVB. However, due to the strong telluric
absorption, we could not use the flux standard in the VIS and NIR; for these wavelengths the continuum shape is uncorrected.

Glass Ia showed characteristics of a K-type stellar source with narrow Ca II H and K emission cores and the Ca II triplet.  It shows weak, symmetric H\textgreek{a} emission with a small self-absorption core, and no Balmer continuum.  The observed spectral features of Glass Ia in the UVB part of the
spectrum are consistent with an early to mid-K type star, as suggested by Chelli et al.~(1988).  We find weak TiO absorption at 615 nm and 705 nm
which would require a low gravity (luminosity class III or lower) at this
spectral type. This in turn is inconsistent with the CO overtone
absorption features at 2300 nm which resemble a late K or early M dwarf. Schmidt et al.
(2012) do in fact resolve Glass Ia as a close double star
in recent epochs of VLT/NACO AO images. Hence, Glass Ia could
consist of an early to mid-K dwarf and an early M dwarf component, which
would explain the spectral features we see.

Glass Ib appears to be a  late G-type star, consistent with the findings of Feigelson \& Kriss (1989).  The most notable emission in Glass Ib is strong H\textgreek{a}, shown in Figure 4, with a blue asymmetry.  The line is wide with FWHM=185 km s$^{-1}$, full width at 10\% of 460 km s$^{-1}$, and equivalent width 24.8 \AA{}.  The H\textgreek{b} emission line, however, is weak, resulting in a very high ratio of H\textgreek{a}/H\textgreek{b}=24. No other Balmer lines were detected.  
While the width and strength of the H\textgreek{a} line suggests active accretion, there are no other accretion indicators.  The spectrum shows no He I emission, except the line at 1083 nm.  The Glass Ib spectrum also does not show the H$_2$ S(1) v=1-2 line at 2.122 \textgreek{m}m, which is used as a tracer of outflows and jets.
There are weak [SII] and [FeII] lines, and stronger, asymmetric [OI] lines, but we do not find higher ionization species.  The [OI] lines show the same velocity profile as H\textgreek{a} but with widths a factor of $\sim$3.5 narrower (see Figure 4).  
The spectrum at longer wavelengths is featureless, likely due to veiling.  
We also find the central star has a high $v$sin$i$ of $\sim$70 km s$^{-1}$, found by modeling the FeI, BaII, and FeII lines at 614 nm because these lines are isolated and dominated by single species.  Nguyen et al.~(2012) also found a high $v$sin$i$ for Glass Ib of 50$\pm$4, but they found the H\textgreek{a} emission in Glass Ib had a full width at 10\% of 318$\pm$20, narrower than found in our X-Shooter spectrum.

\begin{figure}
\begin{center}
\includegraphics[trim=85 80 97 123,clip,angle=90,scale=0.41]{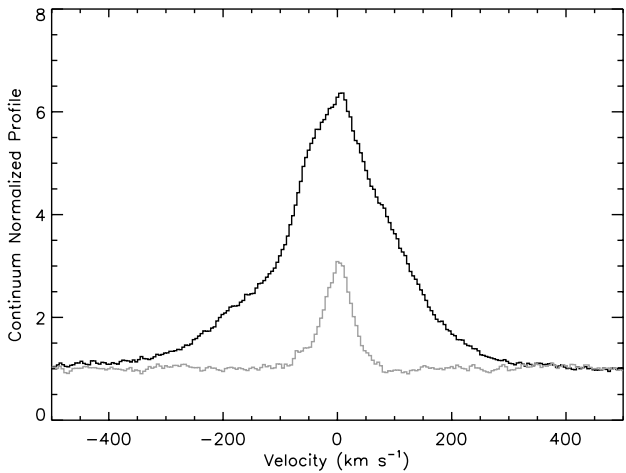}
\caption{The continuum normalized H\textgreek{a} (black spectrum) and [OI] (gray spectrum) emission detected in the X-Shooter spectrum of Glass Ib.}
\end{center}
\end{figure}

We investigated the PSF of the H\textgreek{a}, [OI], and [SII] lines, which are common tracers of outflows, in the raw spectrum.  The plate scale is $0.16^{\prime\prime}\mbox{ pixel}^{-1}$ (22.4 AU assuming a distance of 140 pc) and we found the FWHM of the spatial profile to be 6 pixels ($0.96^{\prime\prime}$ or 134 AU).  By both checking the FWHM in the emission lines and by subtracting the stellar profile, we did not find any extended emission.  We also found the spatial PSF and spatial centroid of these lines are consistent with results at other wavelengths to within 0.1$^{\prime\prime}$, indicating these lines are unresolved.  We also did not see any other extended emission within the 11$^{\prime\prime}$ slit.

\subsection{Optical Photometry}

We also obtained archival imaging data of the Glass I binary system from the FOcal Reducer and low dispersion Spectrograph (FORS; Appenzeller et al.~1998) on the VLT and the
Wild Field Imager (WFI; Baade et al.~1999) at the ESO/MPG 2.2m telescope on La Silla in
various epochs and filters between December 2003 and April 2005. In images
where neither component is saturated, we obtain relative aperture
photometry and measure the brightness ratio of both components of the Glass I binary.  See Table 1 for a detailed list of the archival data used and the measured brightness ratios.

\section{Discussion}

\subsection{The Variable Nature}

The absence of [NeII] emission in the VISIR spectrum is puzzling.  It raises the question of whether the variability is in the photoionizing radiation or the visibility to the ionized region, or both.  Here we investigate the variability of the emission features, extinction, and optical brightness of the Glass I binary in an effort to better understand the source of variability, possibly providing clues to the source of photoionizing radiation.

\subsubsection{Emission}

The Glass I system was also observed with the {\it Spitzer} IRS SH module in March 2006 as part of a survey by Manoj et al.~(2011), and they do not show the fine structure emission lines in their spectrum.  To see if the lines were smoothed out or were weak, we obtained the 2006 SH spectrum from the {\it Spitzer} archive.  We used the post basic calibrated data (pbcd) provided by the archive, removed the fringing using IRSFRINGE, and show the un-smoothed spectrum in Figure 5.  It can be seen there are no traces of the fine structure emission.  We found upper limits for the line intensities, given in Table 2, by measuring the distribution of line strengths at different wavelengths within 1 \textgreek{m}m of each line, and then set the upper limit as the line intensity that would provide a 5-$\sigma$ integrated intensity.  This indicates the fine structure emission is variable as well, which could be due to the ionized gas being hidden (but not the mid-IR continuum emitting region) or a change in the ionization state of the gas due to variation in the ionizing radiation source.  Glassgold et al.~(2007) discussed the timescales for photoionized neon to reach equilibrium in a circumstellar disk, and  found the slowest process related to emission, radiative recombination, is on the order of one year for densities of $\sim10^5$ cm$^{-3}$.  Given that the VISIR spectrum was taken a year and a half after the fine structure emission was detected in the {\it Spitzer} spectrum, we conclude the [NeII] emission weakened below detectable levels.

\begin{figure}
\begin{center}
\includegraphics[trim=85 90 100 115,clip,angle=90,scale=0.41]{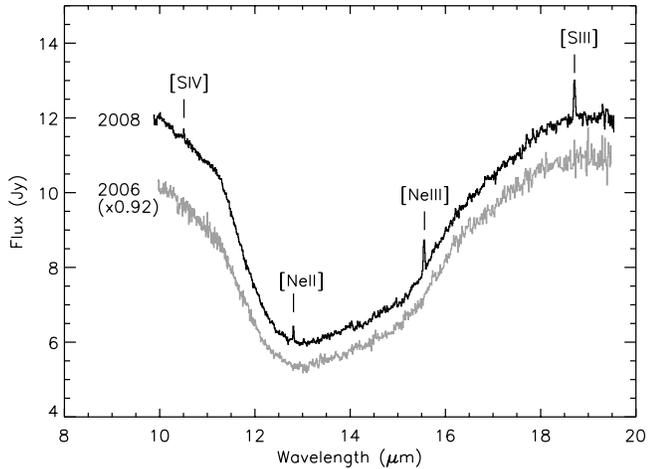}

\caption{The {\it Spitzer} IRS short-high spectrum, dominated by Glass Ib, displaying {[}SIV{]},
{[}NeII{]}, {[}NeIII{]}, and {[}SIII{]} emission as indicated. Overplotted is the {\it Spitzer} IRS spectrum taken in March 2006 as part of a survey by Manoj et al.~(2011).}
\end{center}

\end{figure}

 The absence [NeII] emission in the VISIR spectrum is consistent with the absence of high ionization species in the X-Shooter spectrum taken at roughly the same time.  The X-Shooter spectrum does, however, show H\textgreek{a}, which has proven to be variable as well.  The Glass I system was originally considered a field star because H\textgreek{a} emission was not detected (Glass 1979), but then Feigelson \& Kriss (1989) detected this line toward Glass Ib.  Luhman (2004) detected very little H\textgreek{a} emission, if any, in May 2003, but this feature was very strong in the January 2010 X-Shooter spectrum.  The strength of this line, with an equivalent width of 24.8 \AA, suggests it arises from disk accretion.  If this is the case, this would suggest the accretion onto Glass Ib is changing and at times absent.

The silicate emission in Glass Ib is variable as well.  This was first recognized by G\"urtler et al.~(1999), with the 10 \textgreek{m}m flux decreasing by a factor of 2 between February 1996 and July 1997.  Natta et al.~(2000) observed the system at an intermediate time between those observations in their study of silicate emission from YSOs.  While their models explored the mineralogy of the disks, they acknowledged that their models were unsatisfying in accounting for the changes in silicate emission strength found for Glass Ib.  Other than changes in mineralogy, the variability may be due to changes in dust effective temperature, which depends on the temperature and amount of the dust visible to us.  We find it worth noting that while we show the silicate emission is associated with Glass Ib, Natta et al.~(2000) attributed it to and used the stellar parameters for Glass Ia, which may account for why it showed such a high L$_{sil}/$L$_{\star}$ ratio.

\begin{figure}
\begin{center}
\includegraphics[trim=83 99 97 92,clip,angle=90,scale=0.4]{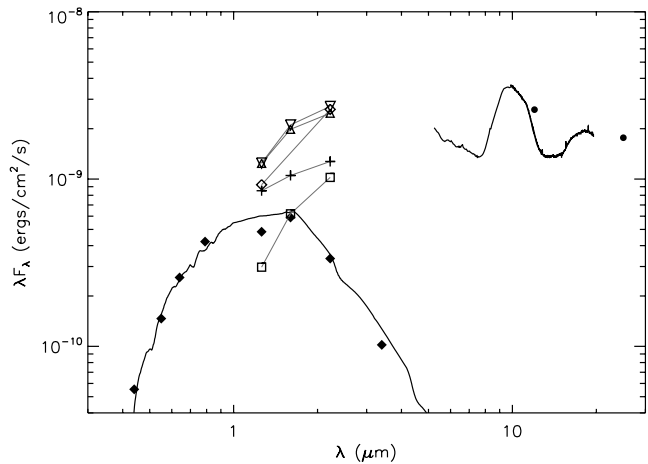}

\caption{A combination of photometry and Spitzer IRS spectra for the Glass I system.
The SED for Glass Ia, taken from Whittet et al.~(1987) and Chelli et al.~(1988),
is shown with solid diamonds.  As mentioned in the text, Glass Ia is actually a binary with K and M star components.  Our model for Glass Ia, seen overplotted, uses stellar spectra with T$_{\star}$=3750, 4500 K and R$_{\star}$=0.6, 0.7 R$_{\odot}$ to represent those respective components, with d=100 pc and A$_{V}$=1.5 mag.  The near-infrared photometric spectra of Glass Ib are shown using open symbols and gray lines, and are taken from Glass (1979; triangles), Chelli et al.~(1988; squares), Cambresy et al.~(1998; diamonds), 2MASS (crosses), and Carpenter et al.~(2002; upside-down triangles) with fluxes from the Glass Ia component, as measured by Chelli et al.~(1988), removed.  The solid circles are the fluxes found with IRAS, and the {\it Spitzer} IRS spectrum is shown, with both being dominated by Glass Ib.  }
\end{center}
\end{figure}

\subsubsection{Extinction}

Glass Ib has proven to be a highly variable object.  Koresko et al.~(1997) noted near- and mid-infrared observations in May 1986 by Chelli et al.~(1988) did not fit well with optical observations in December 1981 by Feigelson \& Kriss (1989), and suggested this may have been due to Glass Ib becoming an infrared object between observations.  Such photometric variability is apparent when comparing the near-infrared observations from the literature.  We provide information from a list of observations in Table 3, and it can be seen the flux varies by almost a magnitude in the {\it JHK}-bands.  In Figure 6, we show these photometric measurements with the Glass Ia flux contribution removed.  To remove the Glass Ia flux, we assume it has remained relatively constant, supported by the fact that the visual brightness was constant at about 12.8 mag during the period 1950-1970 (Feigelson \& Kriss 1989) and still has the same visual magnitude (see section 3.1.3).  It can be seen that while Chelli et al.~(1988) showed Glass Ib dominating the system flux only for wavelengths > 1.6 \textgreek{m}m, all other observations show Glass Ib being brighter than Glass Ia to wavelengths shorter than the J-band.  The near-infrared flux variations are inconsistent with a simple change in the extinction toward the emitting region, which would show a correlation between the brightness and the near-IR colors.  The observed variations more likely reflect a change in the temperature or the distribution of material in our line of sight (or both), as is suggested by variation in the silicate emission described in the previous section.

\begin{figure}
\begin{center}
\includegraphics[trim=85 80 97 105,clip,angle=90,scale=0.4]{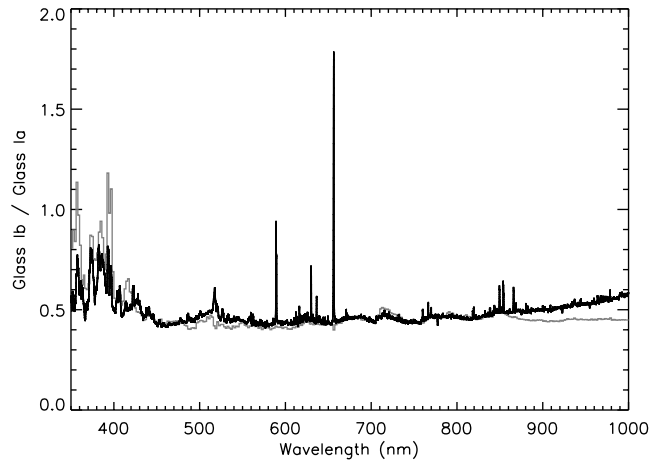}
\caption{The flux ratio of Glass Ib / Glass Ia (black spectrum) as measured with X-Shooter in the UVB and VIS bands.  A 5250 K spectrum is used to model Glass Ib, while 4500 K and 3750 K spectra are combined  to model Glass Ia, with the ratio (gray spectrum) showing the extinction toward Glass Ib is only $\sim$1.3 mag greater than toward Glass Ia. }
\end{center}
\end{figure}


\begin{table}
\begin{center}
\caption{JHK-band observations of the combined Glass I binary components taken from the literature}

\begin{tabular}{ccccc}
\hline 
J & H & K & Obs. Date & Reference\tabularnewline
\hline
\hline 
8.36 & 7.24 & 6.27 & January 1978 & Glass (1979)\tabularnewline
9.22 & 8.06 & 7.06 & May 1986 & Chelli et al.~(1988)\tabularnewline
8.58 & -- & 6.22 & Jan.-May 1996 & Cambresy et al.~(1998)\tabularnewline
8.64 & 7.73 & 6.88 & January 2000 & 2MASS\tabularnewline
8.36 & 7.19 & 6.18 & Jan.-May 2000 & Carpenter et al.~(2002)\tabularnewline
\hline
\end{tabular}
\end{center}
\end{table}

We used the optical X-Shooter spectrum to further investigate the extinction.  As can be seen in Figure 7, the Glass Ib/Ia ratio stays constant at about 0.5 in the 0.5-1.0 \textgreek{m}m range.  For wavelengths longer than 1.29  \textgreek{m}m, Glass Ib increasingly dominates the system flux as expected as dust emission becomes more important.  The ratio in the UV, where dust emission is unimportant, shows the extinction toward Glass Ib is not as high as has been found in the past.  To estimate the extinction in Glass Ib, we used Kurucz (1993) models to model the ratio of Glass Ib/Ia.  We used a 5250 K spectrum to estimate the G-type star in Glass Ib, and divided by a combined 4500 K and 3750 K spectrum to simulate the K-type (Feigelson \& Kriss 1989; Koresko et al.~1997) and M-type (Schmidt et al.~2012) components in Glass Ia.  We then found a good fit with Glass Ib showing only 1.3 mag of visual extinction more than that toward Glass Ia (shown in Figure 7).  Using the visual extinction of 1.7 mag toward Glass Ia found by Koresko et al.~(1997), this would suggest a low A$_{V}$ value of $\sim$3.0 mag for Glass Ib.  This is significantly smaller than the A$_V$ estimate of 17 mag by Koresko et al.~(1997).  This variation in extinction to the central stellar object likely explains the changes in optical brightness of Glass Ib.

\subsubsection{Optical Light Curve}

A careful study of the literature on the Glass I system shows significant variation in the relative optical brightnesses of the two components.  Koresko et al.~(1997) first noted that optical observations from December 1981 did not seem compatible with infrared observations in May 1986, and suggested the system became optically faint between observations.  Luhman (2004) even found that in May 2003, Glass Ib was actually optically brighter than Glass Ia.

To further investigate the optical variability of Glass Ib, we obtained photometric V-band data for the combined Glass I system from the All Sky Automated Survey (ASAS; Pojmanski 2002), shown in Figure 8.  We find the system is variable as expected, with the most striking change being when the system brightened to 10.9 mag  from mid-2002 until mid- to late-2004.  This increased visible brightness is likely due to Glass Ib.  It was during this time that Luhman et al.~(2004) found Glass Ib was the brighter component at 0.66  \textgreek{m}m.  We find Glass Ib was brighter to where it saturated WFI images at 676 nm in December 2003 and at 830 nm in April 2004.  Then in 2005, we find Glass Ib was again fainter than Glass Ia in the U, B, V, and R bands.  We mention these 2005 data simply to provide independent validation for the ASAS data and do not consider the short wavelength bands, which may be influenced by accretion variability, in our discussion of the ASAS time series.  Given that the light curve in Figure 8 never gets fainter than V=12.8 mag, this supports the earlier assumption that Glass Ia is relatively constant at this magnitude and that increases in flux are due to brightness variations in Glass Ib.

\begin{figure}
\begin{center}
\includegraphics[trim=85 80 97 115,clip,angle=90,scale=0.4]{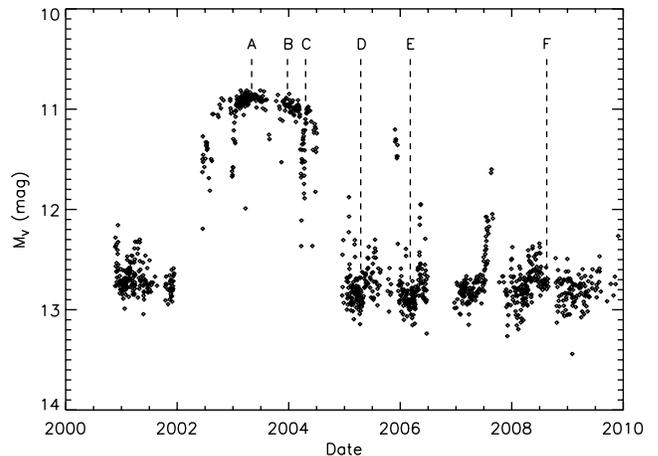}

\caption{The optical light curve of the combined Glass I system from the All Sky Automated Survey (Pojmanski 2002).  The epochs correspond to: A) Luhman 2004 finds Glass Ib is brighter at 0.66  \textgreek{m}m with weak H\textgreek{a} emission, and Meeus et al.~2003 shows no [NeII] or [SIV] in silicate spectrum. B) WFI images at 676 nm show Glass Ib is brighter and saturating image. C) WFI images at 800 nm show Glass Ib is still brighter and saturating image. D) FORS imaging shows Glass Ib is fainter in U, B, V, and R bands. E) {\it Spitzer} IRS observation shows no fine structure emission.  F) {\it Spitzer} IRS observation shows fine structure emission.}
\end{center}
\end{figure}

The increases in visible flux are likely due to variable extinction rather than increased activity.  Referring to Figure 8, it can be seen that during the dark period, there are short-period ($\sim$1-2 weeks) increases in flux.  Then in the bright period, the system is relatively constant with short-period decreases in flux.  This suggests a stellar component in Glass Ib essentially became entirely visible during the bright period, with obscuring material briefly entering the line of sight.  The increases in flux during the dark periods may then be due to variations in the extinction. Also, Luhman et al.~(2004) found the H\textgreek{a} emission toward Glass Ib was small in May 2003 during the bright period, further suggesting this increased brightness is not due to increased activity or accretion.

The length of the bright period ($\sim2.5$ years) and the transition between bright and dark periods ($\sim4-6$ months) are indicate the changes were caused by a thick column that moved out of and into the line of sight.  However, it is difficult to associate these times scales and column density changes with a physical feature in a Keplerian disk, indicating the disk material is most likely not uniform or static.  It may be that the disk is experiencing variable activity such as accretion, as is implied by the variable H\textgreek{a} emission.  If this is the case, the bright period would be when accretion was inactive and the material was out of the line of sight.  

It is noteworthy to point out the optical light curve may show periodicity in the Glass Ib system.  The two large decreases ($\Delta$V$\sim$0.8 mag), centered around January 7, 2003 and April 3, 2004, have similar phase shapes consisting of a flux drop for about 3 weeks, immediately followed by a flux drop for about 2 weeks.  These two phases are shown in Figure 9.  We are unable to identify a periodic trend in the brightness increases during the dark period.  One possible explanation for these flux decreases is that material is eclipsing the central star with a period of 451 days (or possibly half that, the photometric light curve is sparse halfway between the epochs).  This implies the material could be orbiting in the terrestrial planet region, near $\sim$1 AU. However, the periodicity is beyond the scope of this paper, so we leave further investigation and modeling to future work.

\begin{figure}
\begin{center}
\includegraphics[trim=82 85 96 105,clip,angle=90,scale=0.4]{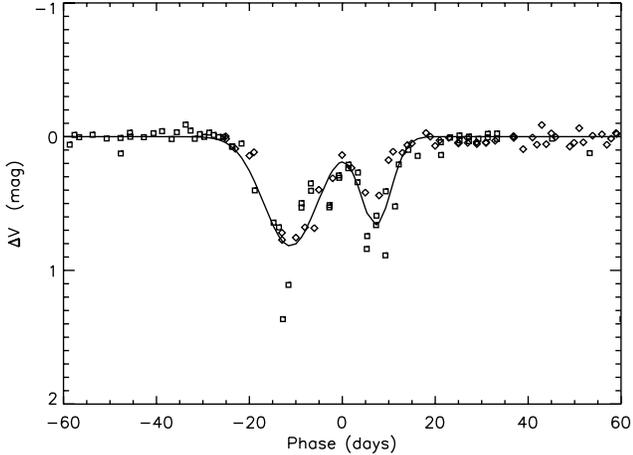}

\caption{Close up of ASAS optical light curve showing possible periodic extinction toward Glass Ib during the "bright period".  Two phases are shown with epochs centered at 2003 January 7 (diamonds) and 2004 April 3 (squares).  The peak fluxes have been subtracted with no correction to the continuum, and a line is drawn to direct the eye to the shape of the flux decreases. }
\end{center}
\end{figure}

\subsection{Nature of the Source}

Glass Ib has been characterized as a young star due to having properties typical of such objects, such as an infrared excess, H\textgreek{a} emission, location in the HR diagram (Chelli et al.~1988), and its association with the Chameleon I (Cha I) star-forming region.  However, strong emission from high-ionization state fine structure lines such as [SIV] and [NeIII] are unusual for young stars and more typical of evolved systems such as planetary nebulae and symbiotic stars.  We consider here some of the evidence for and against each classification.

\subsubsection{Young Star}

While high-ionization state fine structure emission lines are rare in young stars, they have been detected before.  Espaillat et al.~(2012) found a [NeIII]/[NeII] ratio of 1.36 toward the young star SZ Cha, and Green et al.~(2011) detected [ArIII], [NeIII], [OIV], and [NeV] in a Herbig-Haro outflow seen toward GGD 37.  We still investigate if Glass Ib is truly associated with Cha I by first comparing its radial velocity, proper motion, and distance to other pre-main sequence (PMS) stars in the cloud.  If the motion and/or distance to the Glass I system is a strong outlier, it could indicate it did not originate from Cha I. However, if the Glass I binary has a similar motion and distance to the other PMS stars, it would raise doubts that the Glass I system is a field star.

The radial velocities of the young stars in Cha I have a large spread.  Dubath et al.~(1996) used 26 T Tauri stars to find the radial velocities are in the $V_{LSR}\sim2-7$ km s$^{-1}$ range.  We used the isolated FeI and BaII absorption lines near 614 nm in the X-Shooter spectrum to find Glass Ia has a radial velocity of $V_{LSR}=2.2\pm0.7$ km s$^{-1}$, and Glass Ib has a velocity $V_{LSR}=4.8\pm2$ km s$^{-1}$, similar to the other PMS stars.

 For the proper motion of the Glass I system, we show a comparison with the other PMS in Cha I, as well as the nearby field stars, in Figure 10.  The motions for the PMS stars are taken from Teixeira et al.~(2000), and the proper motions for non-PMS field stars within 10$^{\circ}$ are included from Covino et al.~(1997).  Teixeira et al.~(2000) measured the proper motion for the Glass I system as being outside the 2-$\sigma$ of the mean motion for the nearby PMS stars.  However, measurements by both Ducourant et al.~(2005) and Roeser et al.~(2010) found its proper motion is within 1-$\sigma$ of the other PMS stars.

\begin{figure}
\begin{center}
\includegraphics[trim=81 85 95 100,clip,angle=90,scale=0.4]{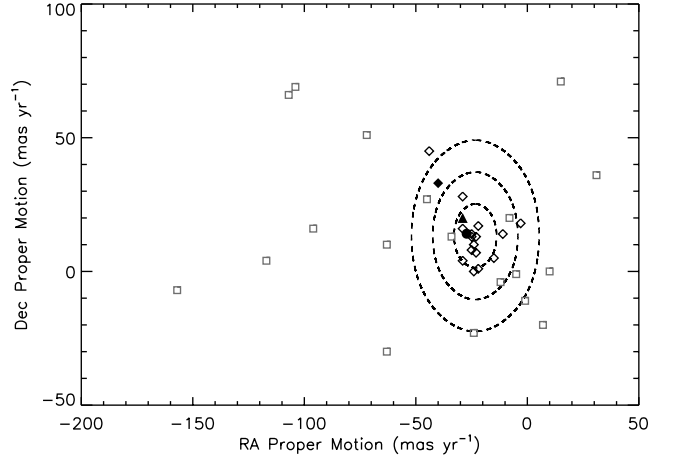}

\caption{Proper motions of the PMS stars (diamonds; Teixeira et al.~2000) and field stars (squares; Covino et al.~1997)
near the Glass I system.   Three proper motion measurements for the Glass I binary are shown with filled symbols taken from Teixeira et al.~(2000; diamond), Ducourant et al.~(2005; triangle), and Roeser et al.~(2010; circle).  The 1-, 2-, and 3-$\sigma$ standard deviations for the PMS proper motions are shown (dashed lines). }
\end{center}
\end{figure}

Previous models of the Glass I system have been fit with an assumed distance of 140 pc, and the Glass I system has been shown to be modeled well as a pre-main-sequence stara at this distance (Koresko et al.~1997).  However, in order to test if the Glass I binary is constrained to be at a similar distance to Cha I, we model Glass Ia as a main sequence star and use that as an estimate for the lower limit on the distance.   We obtained spectra for a 3750 K and a 4500 K stellar spectrum from Kurucz (1993) to estimate the M- and K-type stars in the Glass Ia binary (see section 3.1.2) and fit them to the photometric spectral points from Whittet et al.~(1987) and Chelli et al.~(1988).  Assuming stellar radii of 0.6 and 0.7 R$_{\odot}$, respectively, the minimum distance to the system would be 100 pc (see fit in Figure 6).  Glass Ia has low extinction, suggesting it is not located behind the Cha I star-forming cloud.  As Glass Ia and Ib have common proper motion, they should be at the same distance from the Earth.

We also note that lithium is seen in absorption at 670.8 nm toward both Glass Ia and Ib as expected for YSOs, although this line is also seen in AGB stars.  Except for the high ionization lines, Glass Ib appears to be a young member of Cha I.

\subsubsection{Planetary Nebula}

Planetary nebulae (PN) are old stars that have shed their outer layers in an expanding shell, and the exposed core is hot enough to ionize the surrounding material.  The ratios of the neon and sulfur ionization states are known to be correlated toward emission line objects such as PNe (i.e. Gordon et al.~2008; Hao et al.~2009) with precise values of ratios likely related to the hardness of the underlying radiation (Groves et al.~2008).  To see if Glass Ib has emission expected for a PN, we compare to the ionization ratios in Bernard-Salas et al.~(2008), who used {\it Spitzer} IRS to measure the emission toward PNe.  The {[}NeIII{]}/{[}NeII{]} and {[}SIV{]}/{[}SIII{]} ratios are shown in Figure 11, and Glass Ib is consistent with the linear trend and offset for these objects.  However, the X-Shooter spectrum of Glass Ib does not show any of the ionized emission lines expected from a PN (i.e. OIII, NII, NeIII, etc.) or the H$_2$ emission line at 2.122 \textgreek{m}m that is expected to be seen from the shock regions (van Winckel 2003).

\begin{figure}
\begin{center}
\includegraphics[trim=82 80 85 95,clip,angle=90,scale=0.4]{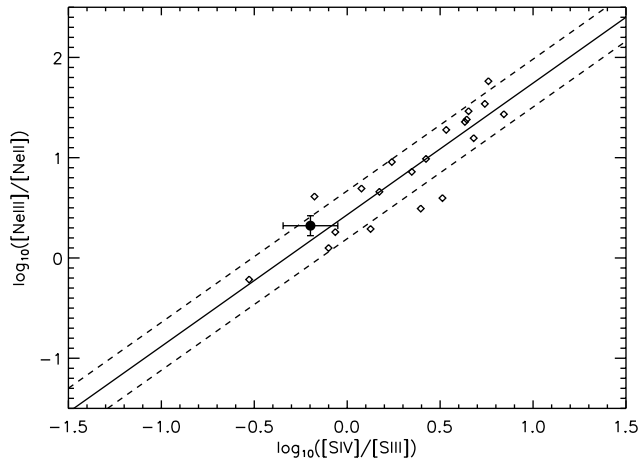}

\caption{Shown is a comparison of ionization abundance ratios in the Glass I binary system (filled
circle) with planetary nebulae showing {[}NeII{]}, {[}NeIII{]}, {[}SIII{]},
and {[}SIV{]} emission as measured in Bernard-Salas et al.~(2008; diamonds).
The correlation between {[}NeIII{]}/{[}NeII{]} and {[}SIV{]}/{[}SIII{]}
is shown with the linear regression (solid) and 1-$\sigma$
error (dashed).  }
\end{center}
\end{figure}

The spectral energy distribution (SED) shape of Glass Ib, shown in Figure 6, suggests that the dust producing the emission is warmer than is typically seen among most PN, possibly indicating the object is a young PN.  The SED of Glass Ib is similar to past observations of the young PN HB 12 (Volk \& Cohen 1990).  However, the [SIV] emission would indicate Glass Ib would be well into the PN stage (K. Volk 2001, \emph{private communication}), as opposed to being a young PN.  Also, the dust in Glass Ib is not likely associated with a shell expanding at a velocity typically seen for a PN.  The silicate in intermediate-mass protoplanetary nebulae is seen in absorption when the shell is optically thick (Garcia-Lario \& Perea Calderon 2003) and again in emission as the object becomes a young PN. As the shell expands further, the peak of the continuum moves to longer wavelengths as the dust cools. If the shell expands on the order of 20-30 km s$^{-1}$, this peak transition from the 10 \textgreek{m}m silicate to longer wavelengths occurs over the period of just a few years (Volk \& Kwok 1989; Volk 1992).   However, the silicate emission has been seen reported toward Glass Ib as early as February 1996 (G\"{u}rtler et al.~1999).   If the dust is in an expanding shell, the continuum would be expected to show stronger infrared excess as opposed to still being similar to IRAS observations (see Figure 6).

While the radiation hardness deducd from the mid-infrared lines is similar to a PN, the behavior of the far-infrared continuum and the [SIV] emission make it unlikely Glass Ib is a young PN, and the strong silicate emission and absence of nebular lines in the optical spectrum make it unlikely Glass Ib is a developed PN.

\subsubsection{Symbiotic Star}

Symbiotic stars (SS) are double star systems where a red giant or Mira variable fills its Roche lobe and transfers material to a nearby white dwarf. A subclass of these objects, called yellow SSs, have spectral types K through A instead of a red giant.  It may be that Glass Ib is a D'-type yellow SS, a subclass of yellow SS that shows strong dust emission at infrared wavelengths (Schmid \& Nussbaumer 1993).  It has been found that D'-type yellow SSs could be a special subclass of SS in that there may have been a mass transfer from the hot component to the cool star instead of vice versa (Pereira et al.~2005).  This results in the cool component having a high {\it v}sin{\it i}, as high as $\sim100$ km s$^{-1}$ (Smith et al.~2001) like that found for Glass Ib.  Glass Ib appears to be similar to the yellow SS V417 Centauri, which was also found to have warm dust, photometric variability, and large variation in the H$\alpha$ emission (van Winckel et al.~1993). 

However, there are major differences between Glass Ib and SSs.  First, the X-Shooter spectrum and all previous optical spectra of Glass Ib do not show ionized nebular emission lines.  The Balmer HI, HeII as well as [OIII], [NeIII], [NeV] and [FeVII] are expected for SSs and are even used as criteria in identifying these objects (Belczynski et al.~2000).  Another major difference is the cold component in SSs have very high luminosities ($10^{2}-10^{4}$ L$_{\odot}$).  If Glass Ib was a SS, it would be located beyond the Cha I star-forming cloud, which again is not likely the case as it is associated with Glass Ia and it does not show high extinction.

\subsection{Ionizing Source}

We consider here the possible origins of such high ionization ratios in a young star.  We investigate if the ionization was caused by an outflow or disk irradiation, and consider the possible mechanisms that may have produced hard radiation.


\subsubsection{Outflow}

One possible source of high ionization is due to the shock temperature in an outflow, which can reach sufficient temperatures of $10^5$ K.  In their theoretical models, Shang et al.~(2010) showed that [NeII] and [NeIII] emission can arise from a jet in a YSO.  They found that higher mass-loss rates and X-ray luminosities result in lower [NeIII]/[NeII] ratios.  Using the line luminosity trends they found for [NeIII] and [NeII], we would expect to have [NeIII]/[NeII]=2.1 with a mass-loss rate $\dot{M}_{\mbox{w}}\approx2\times10^{-9}$ M$_\odot$ yr$^{-1}$ and X-ray luminosity $L_{\mbox{X}}\approx5\times10^{27}$ ergs s$^{-1}$.  High ionization originating in a jet was first detected by Green et al.~(2011), who reported the [ArIII], [NeIII], [OIV], and [NeV] in a Herbig-Haro outflow seen toward GGD 37, although the [NeIII]/[NeII] peaked at $\sim0.2-0.5$ near the outflow source they did not detect [SIII] or [SIV]. 



As indicated above, the lack of [NeII] emission in the VISIR observation may be due to the outflow being extended.  A search in the literature reveals no detections of an outflow associated with Glass Ib.  The nearest Herbig-Haro outflow, HH 935, has a separation of 92$^{\prime\prime}$ north of the Glass I system, but it is moving southward (Bally et al.~2006).  The 2008 {\it Spitzer} IRS spectrum did not show H$_2$ emission at 12.28 and 17.03 \textgreek{m}m, which we would have expected if the fine structure emission was in an outflow.  The other observations of Glass Ib show that it is quiescent compared to expectations for a system with an outlfow.  While Glass Ib shows strong H\textgreek{a} in the X-Shooter spectrum, the H\textgreek{b} line is weak and the other Balmer series lines are absent.  Further, the H\textgreek{a} line is unresolved and originating within 0.1$^{\prime\prime}$ of Glass Ib, so it is difficult to argue that it is originating in an extended outflow.  The X-Shooter spectrum does not show the H$_2$ emission line at 2.122 \textgreek{m}m, a commonly used tracer for outflows and jets.  The Pa\textgreek{b}, Pa\textgreek{g}  and CaII triplet also show strong inverse P-Cygni profiles, which is indicative of infalling matter rather than an outflow.  These factors make it difficult to argue that the fine structure emission is due to the shock of an outflow.

\subsubsection{Disk Irradiation}

We used SIMBAD to search for objects near the Glass I system that may be sources of irradiation, and found that the nearest Herbig-Haro object has a separation of 93$^{\prime\prime}$ from Glass Ib and there are no OB objects within 5$^{\prime}$.  There are no nearby objects that might have irradiated Glass Ib to create the high ionization states, so if the ionization is due to disk irradiation, the radiation is likely originating from the Glass I system.

Stelzer et al.~(2004) detected X-ray emission from the Glass I binary system when probing the Cha I star-forming cloud with {\it XMM-Newton} in April 2002, and the Glass I system had the highest X-ray luminosity of all the sources in their observation.  As disk irradiation likely plays a large part in disk dispersion, models have been created to predict the emission line spectra expected from disks irradiated by X-rays and extreme ultraviolet (EUV) radiation so they can be used as diagnostics for the disk radiation environment (e.g. Ercolano et al.~2008; Meijerink et al.~2008; Ercolano \& Owen 2010).  These models predict lower [NeIII]/[NeII] emission ratios (<0.5) than found toward Glass Ib, more comparable to the ratios previously seen in YSOs.  Glassgold et al.~(2007) showed the [NeIII]/[NeII] ratio that would arise from X-ray ionized neon in equilibrium is consistently small ($\le0.1$) because the X-ray radiation would ionize the gas deeper in the disk and the recombination rate is much higher for Ne$^{2+}$ than Ne$^{+}$. Hollenbach \& Gorti (2009) likewise modeled the X-ray layer as being illuminated with a power-law spectrum $L_{\nu}\propto\nu^{-1}$ for 0.1 keV < $h\nu$ < 2 keV and the [NeIII]/[NeII] ratio was on the order of $\sim0.1$ for all X-ray luminosities up to $\sim10^{31}$ erg s$^{-1}$.  Espaillat et al.~(2012) also found with six T Tauri stars displaying [NeII] and [NeIII] emission that the [NeIII]/[NeII] ratios were smaller for systems showing higher X-ray luminosities.  At first sight, the strong X-ray radiation of log(L$_X$/L$_{bol}$)=-2.65 found toward Glass I by Stelzer et al.~(2004) appears to contradict this trend as it would result in [NeIII]/[NeII] < 0.1 according to the trend found by Espaillat et al.~(2012).  However, the X-ray radiation was observed in April 2002 and Glass I has proven to be highly variable on short time scales, so there may not have been a high X-ray luminosity during the 2008 {\it Spitzer} IRS observation.  If the high ionization in Glass Ib was due to X-ray radiation, the system could not have been in steady state because our [NeIII]/[NeII] ratio is much higher than the models predict (i.e. Glassgold et al.~2007; Hollenbach \& Gorti 2009).  If the ionization was caused by an X-ray flare, it may have been due to a harder spectrum than viewed by Stelzer et al.~(2004) in order to create the strong [SIV] and [NeIII] emission; high energy X-rays are required to eject electrons from the K-shell, and lead to the production of highly ionized species (Glassgold et al.~2007).  As the fine structure emission is so variable, not detected 1.5 years before or after it was seen in the 2008 {\it Spitzer} IRS observation, this speculation seems plausible.

We do note that while a harder ionizing spectrum that creates the high [NeIII]/[NeII] ratio would also be expected to create a higher [SIV]/[SIII] ratio, this does not appear to be the case for Glass Ib.  Models that use a softer spectrum than the power-law spectrum used by Hollenbach \& Gorti (2009) predict higher [SIV]/[SIII] ratios than found for Glass Ib.  For example, Ercolano et al.~(2008) predict a high [SIV]/[SIII]=1.7 while [NeIII]/[NeII]=0.17, and Ercolano \& Owen (2010) predict [SIV]/[SIII]>2 while [NeIII]/[NeII]$\le$0.33 for different X-ray luminosities.  The [SIV]/[SIII] ratio thus appears to be weak for the high [NeIII]/[NeII] ratio found in Glass Ib.  This may be due to the [SIV] being at 10.51 \textgreek{m}m where it is more susceptible to extinction by the silicate feature than the other fine structure emission lines, suggesting the emission is originating in an ionized region that is highly extinguished.

Extreme UV photons may play an important role in the dispersion of disk material, and ultimately in the formation of planets.   Hollenbach \& Gorti (2009) show if the EUV spectrum is sufficiently hard, it can result in high [NeIII]/[NeII] ratios.  If we assume Glass Ib is at a distance of 140 pc, the [NeIII] line luminosity would be $4.6\times10^{-5}$ L$_{\odot}$.  This high line intensity can be produced if the disk is illuminated by an EUV photon luminosity of $\ge10^{43}$ s$^{-1}$, which would also create a high [NeIII]/[NeII] ratio $\ge$2 (cf. Figure 1; Hollenbach \& Gorti 2009).  These ratios are created when $\nu L_{\nu}$ is constant from 13.6-100 eV, which Hollenbach \& Gorti (2009) estimated from young stars typically having this continuum shape in their far-UV and X-ray spectra.  However, as the [NeIII]/[NeII] ratios previously detected in YSOs are much smaller, they instead implemented a softer EUV spectrum for their models.  Our detection of higher ionization ratios may indicated a power-law spectrum for this object.

Whether the ionization is caused by X-ray or EUV radiation, the source of the radiation remains a mystery.  Two possible sources include an accretion shock or strong magnetic activity.   Although variability is regularly seen in magnetically active stars, we also consider whether variable accretion could lead to the photoionizations.  The silicate emission and nIR continuum indicate changes in extinction and the effective temperature of the dust.  The optical light curve implies the distribution of material in the disk is irregular and changes over time, and the H\textgreek{a} emission indicates the presence of variable activity in the system.  These different factors together suggest the system could be experiencing variable accretion.  Hollenbach \& Gorti (2009) show that in order for EUV photons to reach the disk beyond 1 AU, the disk wind has to have a mass loss rate $\le10^{-9}$ M$_\odot$ yr$^{-1}$, corresponding to an accretion rate of $\le10^{-8}$ M$_\odot$ yr$^{-1}$.  If the accretion rate decreased, it may be that the EUV and soft X-ray radiation escaped out the "sides" of the accreting columns (Drake 2005; Ardila 2007; Ardila \& Johns-Krull 2009) while the central star was still producing the photoionizing radiation.  A special circumstance and timing such as this would explain why fine structure emission was not detected toward Glass Ib before or after our {\it Spitzer} IRS observation.

It is unclear if this photoionizing event is due to a special circumstance that is specific to Glass Ib, or if such an event also occurs in other young stars.  It may be that the EUV radiation that is predicted to illuminate disks is infrequent and short-lived, which could explain why highly ionized fine structure emission has not been seen in other disks.  Further studies of the Glass I binary system are necessary to monitor the variability and investigate the production of photoionizing radiation, allowing us to understand if this event could be common in other disks as well.

\section{Summary}

Glass Ib has proven to be a highly variable system.  We found a low visual extinction in the system of A$_{V}$=3.1 mag, much lower than the previous estimate of 17 mag by Koresko et al.~(1997).  The optical light curve showed Glass Ib experienced a period of little extinction toward the central emitting source, making it temporarily brighter than Glass Ia.  The length of time of the bright period, and the sharp transitions between dark and bright periods, suggest the extinguishing material is not in a stable hydrostatic disk, but is rather variable and dynamic.  The combination of variable silicate emission and history of the near-infrared photometry further indicate variable extinction to the central star and changes in the effective temperature of the disk.  Finally, the H\textgreek{a} emission indicates variable activity in the system.  These different properties could indicate the system is undergoing variable accretion.

The ionization ratios found ([NeIII]/[NeII]=2.1 and [SIV]/[SIII]=0.6) are much higher than has ever been detected in a young star, more typical of that found in a planetary nebula.  To test the characterization of Glass Ib as a young star, we investigated the proper motions and modeled the spectral energy distributions of Glass Ia and Ib and found the system is most likely associated with the Chameleon I star-forming region.  Our models fit well with the SEDs using a G-type star for Glass Ib, and a combined K-type and M-type to simulate a binary in Glass Ia (Schmidt et al.~2012).  Glass Ib also shows HCN, CO$_2$, and H$_2$O emission, which are commonly found in young stars (Pontoppidan et al.~2010).  

We investigated whether the fine structure emission was due to an outflow or from the disk being irradiated by hard photons.  While the shock fronts of outflows can produce high ionization states, we were unable to find any indicators that an outflow is present in Glass Ib.  We found Glass Ib does not display the H$_2$ emission line at 2.122 \textgreek{m}m and, aside from H\textgreek{a}, the Balmer series lines are weak or absent.  Further, the Paschen series and CaII triplet appear to instead suggest an inflow with strong inverse P-Cygni profiles.

If the emission is due to hard radiation, the source of the radiation is unclear, possibly arising from  magnetic activity or an accretion shock.  In either case, if the ionization was due to EUV irradiation, it was likely a power-law spectrum as this would produce a high [NeIII]/[NeII] ratio in equilibrium (Hollenbach \& Gorti 2009).  If the high ionization was due to hard X-ray irradiation, we likely did not see the system in equilibrium as this would produce a much lower [NeIII]/[NeII] ratio.  It may be that we observed the disk being irradiated by an X-ray flare, or X-rays originating from an accretion event.  A special circumstance and timing, such as observing an X-ray flare, can potentially explain why highly ionized fine structure emission was not seen toward Glass Ib in other observations, and possibly why it has not been seen in other young stars.  While models predict that hard radiation can produce the high [NeIII]/[NeII] ratios, a higher [SIV]/[SIII] would be expected (Ercolano et al. 2008; Ercolano \& Owen 2010).  The weaker [SIV]/[SIII] ratio in Glass Ib may be due to the [SIV] being at 10.51 \textgreek{m}m where it is obscured by silicate absorption, suggesting the fine structure emission arises from an extinguished region.

\acknowledgements{

We would like to thank Serge Correia, Giles Novak, Kevin Volk, John Lacy and the anonymous reviewer
for their constructive help, comments, and suggestions.
Support for this work was provided by the National Science Foundation under Grant
No. AST-0708074, and by NASA through contract RSA No. 1346810, issued
by JPL/Caltech. This work is based on observations made with the {\it Spitzer} Space Telescope, the Gemini Observatory, and the 
ESO Telescope at the Paranal Observatory under programme IDs 072.A-9005(A), 072.A-9016(A), 075.C-0335(A), and 084.C-1095(A).
The {\it Spitzer} Space Telescope is operated by the Jet Propulsion Laboratory,
California Institute of Technology under a contract with NASA. The
Gemini Observatory is operated by the Association of Universities
for Research in Astronomy, Inc., under a cooperative agreement with
the NSF on behalf of the Gemini partnership: the National Science
Foundation (United States), the Science and Technology Facilities
Council (United Kingdom), the National Research Council (Canada),
CONICYT (Chile), the Australian Research Council (Australia), Minist\'{e}rio
da Ci\^{e}ncia e Tecnologia (Brazil) and Ministerio de Ciencia, Tecnolog\'{i}a
e Innovaci\'{o}n Productiva (Argentina). This paper is also made use of the SIMBAD database, operated at CDS, Strasbourg, France.  Basic research in infrared 
astronomy at the Naval Research Laboratory is supported by 6.1 base funding.}

\end{document}